\documentclass[%
 aip,
 jmp,%
 amsmath,amssymb,
 reprint,%
]{revtex4-1}

\usepackage{graphicx}
\graphicspath{{figures_pdf/}}
\usepackage{dcolumn}
\usepackage{bm}
\usepackage{subfigure}

\usepackage{soul}
\usepackage{amsmath,amssymb,amsfonts}
\usepackage{graphics}
\usepackage{color}
\usepackage{xcolor}
\definecolor{b}{rgb}{0,0,1}

\usepackage{mathtools, nccmath}
\usepackage{algorithm}
\usepackage{algorithmic}
\usepackage{enumitem}
\setlist[itemize]{leftmargin=*}

\usepackage{comment}
\usepackage{hyperref}
\usepackage{lipsum}
\usepackage{tabularx}
\usepackage{mathrsfs}
\usepackage{stackengine}

\begin{document}








\title{Interface learning in fluid dynamics: statistical inference of closures within micro-macro coupling models}




\author{Suraj Pawar}
\author{Shady E. Ahmed}

\author{Omer San}%
 \email{osan@okstate.edu}
 

\affiliation{ 
School of Mechanical \& Aerospace Engineering, Oklahoma State University, Stillwater, OK 74078, USA.
}%



\date{\today}

\begin{abstract}
Many complex multiphysics systems in fluid dynamics involve using solvers with varied levels of approximations in different regions of the computational domain to resolve multiple spatiotemporal scales present in the flow. The accuracy of the solution is governed by how the information is exchanged between these solvers at the interface and several methods have been devised for such coupling problems. In this article, we construct a data-driven model by spatially coupling a microscale lattice Boltzmann method (LBM) solver and macroscale finite difference method (FDM) solver for reaction-diffusion systems. The coupling between the micro-macro solvers has one to many mapping at the interface leading to the interface closure problem, and we propose a statistical inference method based on neural networks to learn this closure relation. The performance of the proposed framework in a bifidelity setting partitioned between the FDM and LBM domain shows its promise for complex systems where analytical relations between micro-macro solvers are not available.
 
\end{abstract}


\keywords{Interface closure, Machine learning, Micro-macro coupling, Multiscale systems} 
\maketitle

\emph{Introduction} --- Many problems in science and engineering are multiscale in nature consisting of different spatial and temporal scales that interact with each other \cite{weinan2007heterogeneous,koumoutsakos2005multiscale,fish2010multiscale,matouvs2017review}. For example, simulation of turbulent and separated flows requires resolving a wide range of interacting scales \cite{sagaut2013multiscale,koumoutsakos2005multiscale}. The traditional numerical setup usually focuses on the macroscale evolution of the system and either eliminates meso/microscale processes or consider them through some form of closure models \cite{van1992stochastic,weinan2007heterogeneous}. For many complex systems, modeling coarse-scale, macroscopic behavior might be impractical without simplified assumptions, and one might have to resort to modeling fine-scale, microscale processes to get accurate solution \cite{keunings2004micro,attinger2003generalized,kevrekidis2009equation}. However, using the microscale solver over the entire computational domain is computationally prohibitive even in the exascale computing era. This difficulty can be overcome by utilizing a micro-macro coupled solver to recover accurate solutions in a computationally efficient manner.        

There have been several studies that deal with coupling micro-macro solvers, such as finite difference method-lattice Boltzmann method (FDM-LBM) solver \cite{albuquerque2004coupling,van2007accuracy}, finite volume method-molecular dynamics (FVM-MD) solver \cite{nie2004continuum}, finite volume method-lattice Boltzmann method (FVM-LBM) solver \cite{mishra2007solving,joshi2010hybrid,luan2011evaluation}, molecular dynamics-lattice Boltzmann (MD-LBM) solver \cite{dupuis2007coupling}. These approaches can also be interpreted in a broad sense as domain decomposition methods, where two solvers simultaneously advance the multiscale and multiphysics problems and the information is exchanged across the interface between solvers \cite{chan1989domain,tang2020review}. One of the major challenges in these micro-macro solvers is a mismatch in the kind and number of variables used by different models. For example, microscopic solvers such as LBM describe the evolution of the system through particle distribution functions that are restricted to move on a grid with certain velocities only. On the other hand, the macroscale solvers like partial differential equations (PDEs) model the system in terms of observables like flow velocity, pressure, and density. The coupling between two solvers should not produce any discontinuity in the interface region. Also, the global quality of the numerical solution will be dependent on the treatment of interface boundary condition. Therefore, the interface boundary closure is a major challenge in modeling and computation of these micro-macro solvers. Some of the methods to determine interface boundary closure are based on physical quantities of another solver or from an analytical relation between micro and macro solvers \cite{albuquerque2004coupling,van2006numerical}. However, the derivation of such analytical expressions for complex geometries or systems with complex local interactions might be impractical and for these systems iterative numerical algorithms have been proposed to close interface boundary conditions \cite{van2007accuracy,gear2005projecting}.

The challenges in the treatment of interface boundary conditions in complex systems offer opportunities for researchers to develop statistical inference approaches for coupling micro-macro solvers. In fact, several data-driven approaches have been proposed in the literature to learn the correlation between coarse-scale PDEs and fine-scale microscopic processes \cite{sirisup2005equation,kevrekidis2003equation,siettos2012equation,lee2020coarse}. The machine learning (ML) algorithms are also proven to be successful in discovering hidden PDEs from macroscopic observations data \cite{raissi2018hidden,rudy2017data,long2018pde}. The recent advancement in the efficient implementation of ML algorithms along with the huge amount of archival data generated from high-fidelity numerical simulations and experiments has made statistical inference approaches an attractive choice for multiscale systems. To this end, we propose a data-driven ML-based interface closure framework for the spatial coupling of microscale LBM and macroscale FDM solvers. Our framework is solver agnostic in the sense that it can be extended to any type of micro-macro coupled models. We illustrate our framework for a reaction-diffusion system (the FitzHugh-Nagumo model) and also compare our results against two methods for solving interface closure problems based on numerical approximations of the analytical relations.

\emph{FitzHugh-Nagumo model} --- We illustrate the statistical interface closure for micro-macro coupled solvers using the FitzHugh-Nagumo model. This model consists of two reactive-diffusion partial differential equations, whose dynamics is governed by 
\begin{align}
    \frac{\partial u}{\partial t} &= D^u \frac{\partial ^2 u}{\partial x^2} + u - u^3 -v, \label{eq:fn_model_u} \\ 
    \frac{\partial v}{\partial t} &= D^v \frac{\partial ^2 v}{\partial x^2} + \epsilon(u - a_1 v - a_0),\label{eq:fn_model_v}
\end{align}
where $D^u$ and $D^v$ are the diffusion coefficients of $u$ and $v$, $a_1$ and $a_0$ are model parameters, and $\epsilon$ represents a kinetic bifurcation parameter. We set the parameters to $a_1=2, a_0=-0.03, \epsilon=0.01, D^u=1,$ and $D^v=4$ as suggested by \citet{theodoropoulos2000coarse}. Our spatial domain extends between $[0,20]$ and the model is integrated from time $t=0$ to $t=450$.

We employ FDM as a macro solver and LBM as the micro solver. For the FDM solver, Equation~\ref{eq:fn_model_u} and Equation~\ref{eq:fn_model_v} are discretized with explicit forward in time and the central difference in space as follow 
\begin{equation}
\begin{multlined}
{u}(x,t+\Delta t) = u(x,t) + D^u \frac{\Delta t}{\Delta x^2} (u(x+\Delta x,t) -  \\ 2 u(x,t) + u(x-\Delta x,t)) + \Delta t (u - u^3 -v) , 
\end{multlined}
\end{equation}
\begin{equation}
\begin{multlined}
{v}(x,t+\Delta t) = v(x,t) + D^v \frac{\Delta t}{\Delta x^2} (v(x+\Delta x,t) -  \\ 2 v(x,t) + v(x-\Delta x,t)) + \epsilon \Delta t (u - a_1 v - a_0) , 
\end{multlined}  
\end{equation}
where $\Delta x$ and $\Delta t$ are the spatial and temporal discretization steps, respectively. Since the above discretization is explicit in time, the time step in the above scheme is restricted by the stability condition $\Delta t < C \Delta x^2$. In the FitzHugh-Nagumo model, $D^v > D^u$, and therefore the stability condition will be governed by the diffusion constant of the $v$ equation, i.e., $\Delta t < \Delta x^2/(2 D^v)$.  

The LBM method used as a micro solver describes the evolution of particle distribution functions ${f}_i(x,t)$ discretized in space $x$, and time $t$ along the $i^{\text{th}}$ direction with velocity $c_i$ \cite{chen1998lattice}. We utilize the D1Q3 model, where D1 denotes the one-dimensional domain, and Q3 stands for three velocities. The lattice Boltzmann equation (LBE) describing the evolution of particle distribution is given as 
\begin{align}
\begin{multlined}
    {f}_i^l(x+c_i\Delta x,t+\Delta t) = {f}_i^l(x,t) - \\ {\omega^l} ({f}_i^l(x,t) - {f}_i^{l,\text{eq}}(x,t)) + {R}_i^l(x,t), \quad l \in \{u,v\}
\end{multlined} \label{eq:particle_distribution}
\end{align}
for $c_i=i \in \{-1,0,1\}$. The first step of the LBM is the collision step, where this diffusive collisions are modeled by the Bhatnagar–Gross–Krook (BGK) model as a relaxation to the equilibrium particle distribution ${f}_i^{l,\text{eq}}(x,t)$ with a relaxation coefficient ${\omega^l}$ (denoted by the second term in Equation~\ref{eq:particle_distribution}), and the reactions are modeled by the term ${R}_i^l(x,t)$ \cite{ponce1993lattice}. The second step of the LBM is the propagation of the particle distribution functions to a neighboring grid based on the $i^{\text{th}}$ direction as shown on the left hand side of Equation~\ref{eq:particle_distribution}. The equilibrium particle distribution function, relaxation coefficients, and the reaction term are computed as below
\begin{align}
{f}_i^{u,\text{eq}}(x,t) &= \frac{1}{3}{u}(x,t), \quad {f}_i^{v,\text{eq}}(x,t) = \frac{1}{3}{v}(x,t), \\ 
{\omega^u} &= \frac{2}{1 +3 {D^u} \frac{\Delta t}{\Delta x^2}}, \quad {\omega^v}= \frac{2}{1 +3 {D^v} \frac{\Delta t}{\Delta x^2}}, \\
{R}_i^u(x,t) &= \frac{(u - u^3 -v)}{3}, {R}_i^v(x,t) = \frac{\epsilon (u - a_1 v - a_0)}{3}  .
\end{align}
The observable $u$ and $v$ are defined as the zeroth order moment of the distribution functions 
\begin{equation}
    {u} = \sum_{i=-1}^{i=1} {f}_i^u (x,t), \quad {v} = \sum_{i=-1}^{i=1} {f}_i^v (x,t). \label{eq:lbm_uv}
\end{equation}

\emph{Coupling between FDM and LBM} --- In order demonstrate the coupling between macroscale FDM solver and microscale LBM solver, we divide our computational domain into two non-overlapping subdomains as shown in Figure~\ref{fig:interface}. The left domain is solved using FDM and the right domain is resolved with LBM. The interface lie at the center of the domain, i.e., at $L/2$, where $L$ is the length of the whole domain. Therefore, the last grid point of the FDM is located at $L/2-\Delta x/2$ and the first point of LBM is at $L/2+\Delta x/2$. For our analysis, we assume that the spatial discretization $\Delta x$ and time step $\Delta t$ for both subdomains are the same. When the hybrid solver is used to model the dynamics of reactive-diffusion system, the information has to be exchanged carefully across the interface as the FDM and LBM use different sets of variables, namely $(u,v)$ versus $(f_i^u,f_i^v)$. In order to solve the FDM equation at the last point of the left domain, we need information of $(u,v)$ at $x_{\text{FDM}}+\Delta x$ location which can be obtained using the relation provided in Equation~\ref{eq:lbm_uv} as follow
\begin{align}
    {u}(x_{\text{FDM}}+\Delta x,t) &= \sum_{i=-1}^{i=1} {f}_i^u (x_{\text{LBM}},t),\\  {v}(x_{\text{FDM}}+\Delta x,t) &= \sum_{i=-1}^{i=1} {f}_i^v (x_{\text{LBM}},t). \label{eq:fdm_ghost}
\end{align}

\textbf{\begin{figure}[ht]
\centering
\includegraphics[width=0.95\linewidth]{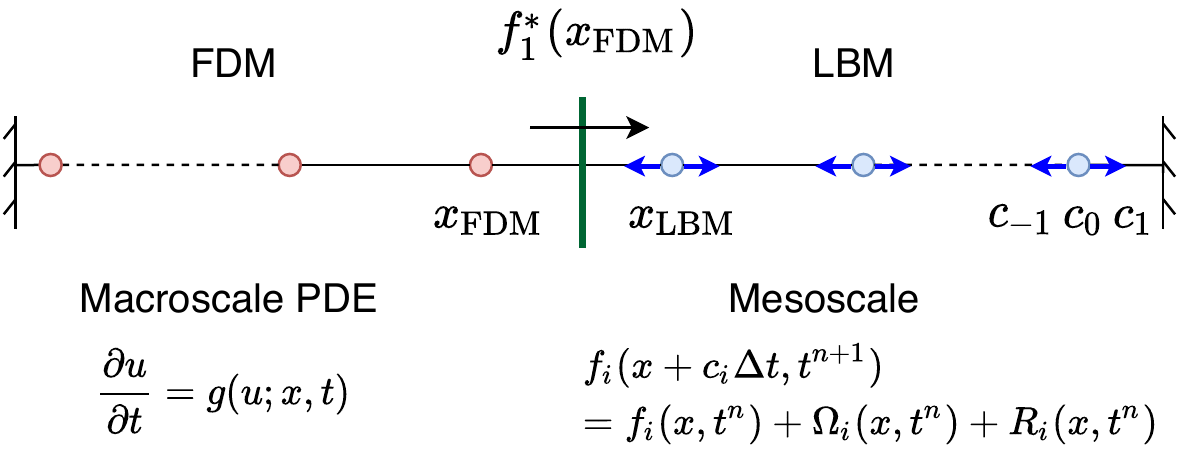}
\caption{The exchange of information across the interface between two domains. The left domain is solved using FDM and the right domain is solved with LBM and the information is exchanged with an appropriate boundary condition.}
\label{fig:interface}
\end{figure}
}

The inverse mappings from FDM to LBM, i.e., $u(x_{\text{FDM}},t) \rightarrow f_i^u(x_{\text{FDM}},t)$ and $v(x_{\text{FDM}},t) \rightarrow f_i^v(x_{\text{FDM}},t)$ for $i \in \{-1,0,1\}$ are not straightforward. We need $f_1^l(x_{\text{FDM}},t)$ to evolve particle distribution at $x_{\text{LBM}}$ from time $t$ to $t+\Delta t$ using Equation~\ref{eq:particle_distribution}. The particle distribution in LBM can be reduced in terms of the FDM variables using a multiscale Chapman–Enskog expansion \cite{chen1998lattice,chopard2002cellular,van2006numerical} as follow
\begin{equation}
    f_i^l(x,t) = f_i^{l,[0]}(x,t)  + f_i^{l,[1]}(x,t)  + f_i^{l,[2]}(x,t)  + \dots, \label{eq:ce}
\end{equation}
where the superscript inside a square bracket denotes the order of approximations. 

We can reconstruct $f_1^l(x_{\text{FDM}},t)$ using the zeroth order Chapman–Enskog relations(CE-0) as shown below
\begin{align}
    f_1^u(x_{\text{FDM}},t) = f_1^{u,[0]}(x_{\text{FDM}},t) = \frac{u(x_{\text{FDM}},t)}{3}, \\
    f_1^v(x_{\text{FDM}},t) = f_1^{v,[0]}(x_{\text{FDM}},t) = \frac{v(x_{\text{FDM}},t)}{3}. 
\end{align}
The above relation is the simplest approximation one can use and the relation given in Equation~\ref{eq:ce} describes the particle distribution in LBM up to the second order. Once the particle distribution function is reconstructed, it has to be propagated from time $t$ to $t+\Delta t$ in two steps. First, the collision and reaction go from time $t$ to $t^*$, and then the post-collision distribution $f_i^{l,*}$ is propagated from time $t^*$ to $t+\Delta t$. The post-collision particle distributions are computed as follow
\begin{align}
\begin{multlined}
    f_1^{u,*}(x_{\text{FDM}},t) = (1-\omega^u) f_1^u(x_{\text{FDM}},t) + \\  \frac{\omega^u}{3}u(x_{\text{FDM}},t) + \frac{\Delta t}{3}{R}_i^u(x,t)
\end{multlined},
\end{align}
\begin{align}
\begin{multlined}
    f_1^{v,*}(x_{\text{FDM}},t) = (1-\omega^v) f_1^v(x_{\text{FDM}},t) + \\  \frac{\omega^v}{3}v(x_{\text{FDM}},t) + \frac{\Delta t}{3}{R}_i^v(x,t)
\end{multlined}. 
\end{align}
Finally, the post-collision particle distributions are propagated to $x_{\text{LBM}}$ as given below
\begin{align}
    f_1^{u}(x_{\text{LBM}},t+\Delta t) &= f_1^{u,*}(x_{\text{FDM}},t), \\
    f_1^{v}(x_{\text{LBM}},t+\Delta t) &= f_1^{v,*}(x_{\text{FDM}},t)
\end{align}
The analytical relation provided in Equation~\ref{eq:ce} might not always be available or might be difficult to derive numerically due to complex geometry or complex local interaction force terms. There are iterative numerical algorithms like constrained runs (CR) scheme \cite{gear2005projecting,gear2005constraint} that can be applied to approximate particle distribution numerically. The correlation between microscale and macroscale variables can also be learned with data-driven methods and we explore the feasibility of feedforward neural network to approximate these relations.  We train the neural network to learn the particle distribution $f_1^l(x_{\text{LBM}},t)$ at time $t$ based on the local information at the previous time step. More concretely, our neural network is trained to learn the below mapping
\begin{equation}
\begin{multlined}
\{u(x_{\text{FDM}},t-\Delta t), v(x_{\text{FDM}},t-\Delta t), f_{-1}^u(x_{\text{LBM}},t-\Delta t), \\ f_{-1}^v(x_{\text{LBM}},t-\Delta t) \} \rightarrow \{f_{1}^u(x_{\text{LBM}},t),  f_{1}^v(x_{\text{LBM}},t) \}.        
\end{multlined}
\end{equation}
The training data for the neural network is generated by simulating the whole computational domain using FDM and LBM separately. We employ bounce-back boundary condition on left and right side of the domain for LBM simulation. The initial condition for the training data is generated with $\alpha=0.7$ as follow
\begin{align}
    u(x,0) &= \alpha ~\text{tanh}(x-L/2), \label{eq:u0} \\
    v(x,0) &=  0.1\bigg(1 + \text{tanh}\bigg(\frac{x-L/2}{4}\bigg)\bigg).
\end{align}
During testing, we use $\alpha=1.0$ so that there is no overlap between training and testing data. We employ a fairly simple neural network architecture with a single hidden layer consisting of eight neurons to learn the correlation between inputs and outputs. More sophisticated architecture can also be employed for complex geometries or high-dimensional problems with complex local interactions.   

\begin{figure}[ht]
\centering
\includegraphics[width=0.99\linewidth]{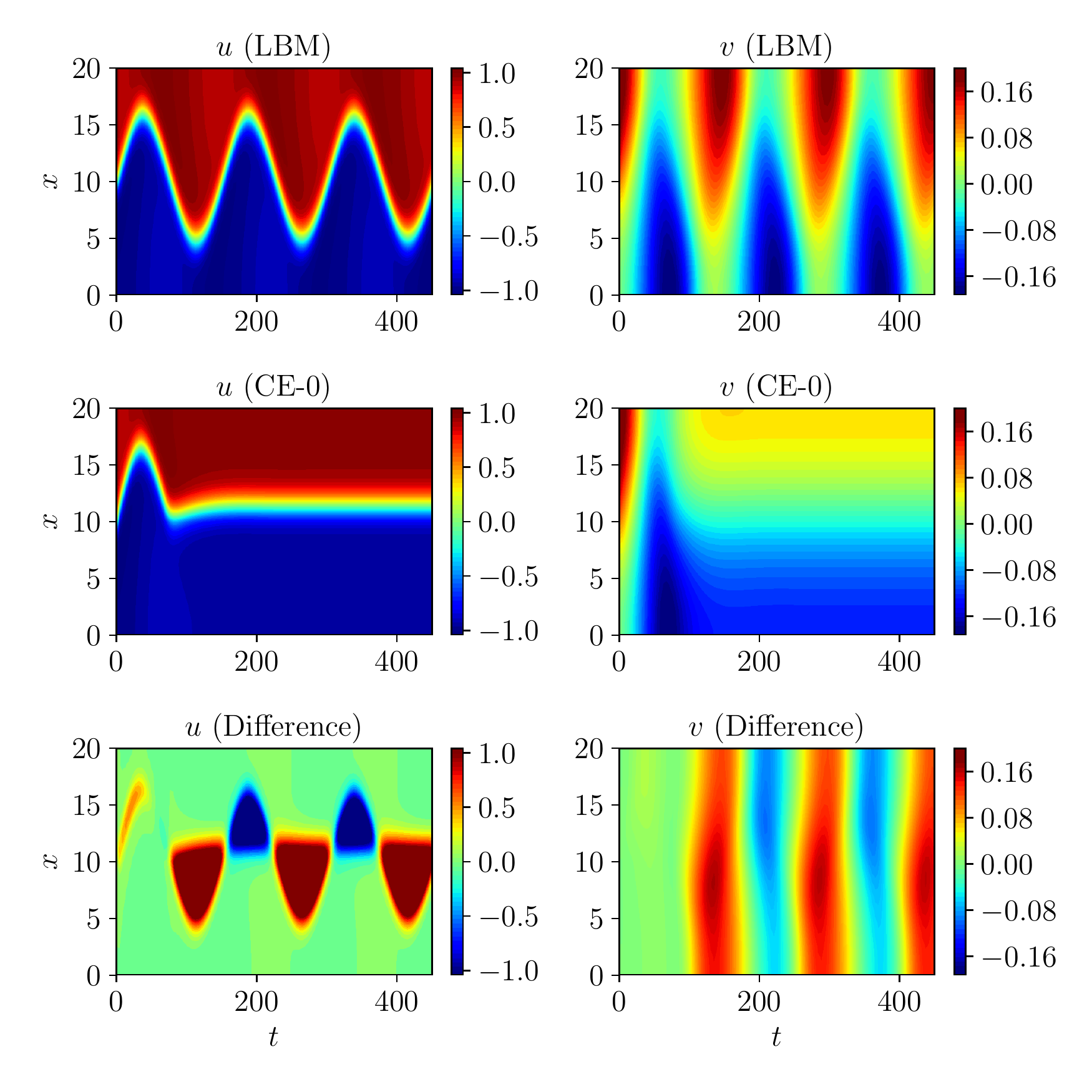}
\caption{Spatio-temporal prediction of the $u$ and $v$ field of the FitzHugh-Nagumo model by hybrid solver with CE-0 coupling at the interface.}
\label{fig:ce0}
\end{figure}

\emph{Results and discussion} --- Here, we present the results of coupled FDM-LBM solver for the FitzHugh-Nagumo model and compare it against the LBM simulation. The computational domain is discretized with a spatial grid of size $\Delta x = 0.1$ and the system is evolved in time with a time step of $\Delta t = 0.001$ from time $t=0$ to $t=450$. The results are provided for $\alpha = 1.0$ in the initial condition of $u$ as given in Equation~\ref{eq:u0}. Figure~\ref{fig:ce0} shows the plot of $u$ and $v$ for the LBM simulation, the FDM-LBM solver with CE-0 coupling at the interface, and the difference between two simulations. It can be seen that the CE-0 coupling is not able to predict the accurate dynamics after time $t \approx 80$. The local error for CE-0 coupling at the interface depends upon the dominant term left out in the Chapman–Enskog expansion given in Equation~\ref{eq:ce}. Therefore, the CE-0 introduces an error which is first order in $\Delta x$. This error seems to have contaminated the solution and the coupled solver based on CE-0 leads to the poor prediction of the $u$ and $v$ fields. 

\begin{figure}[ht]
\centering
\includegraphics[width=0.99\linewidth]{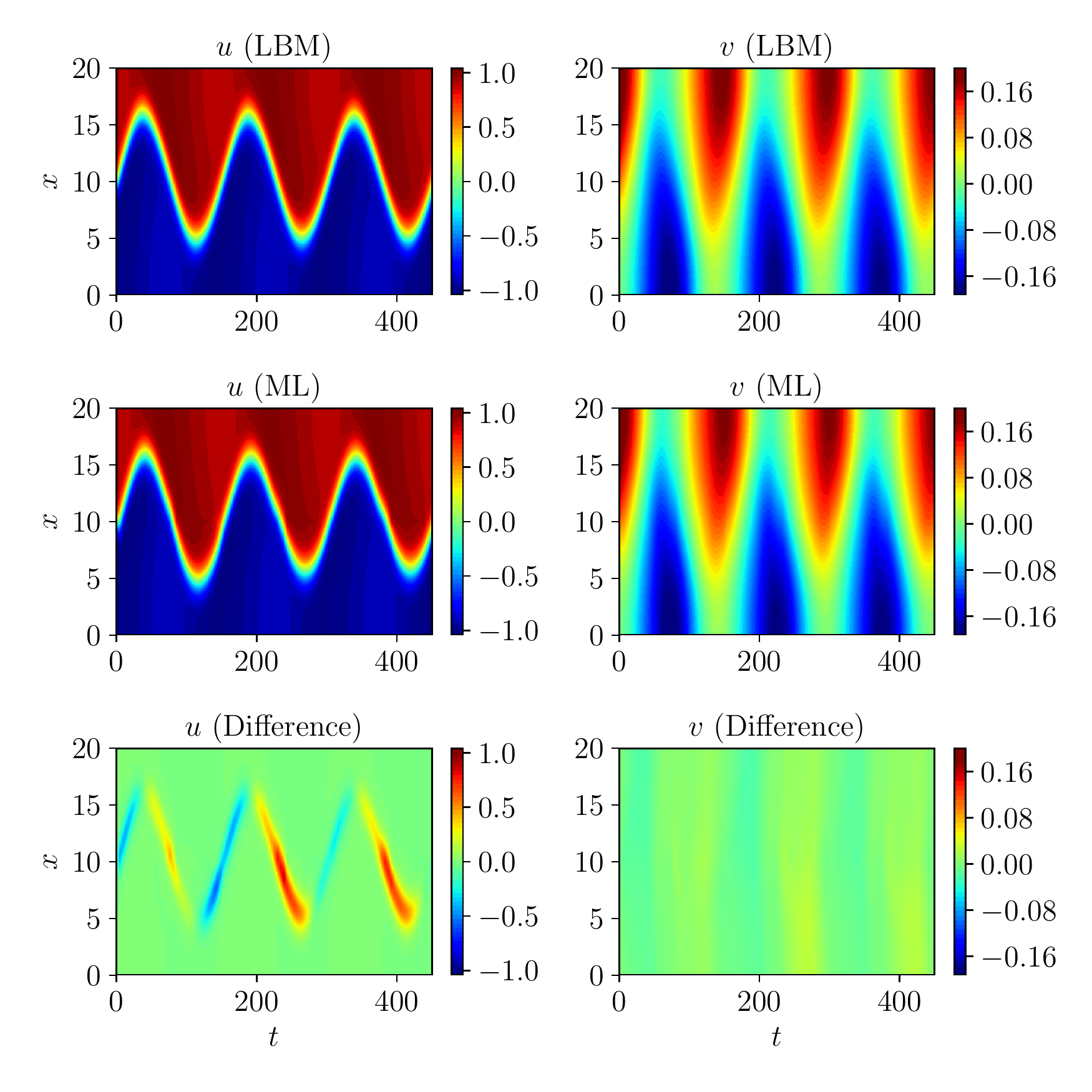}
\caption{Spatio-temporal prediction of the $u$ and $v$ field of the FitzHugh-Nagumo model by hybrid solver with ML based coupling at the interface.}
\label{fig:ml}
\end{figure}

In Figure~\ref{fig:ml}, we depict the results for coupled FDM-LBM solver using ML-based coupling at the interface. The ML-based coupling can predict the correct dynamics of both $u$ and $v$ fields with a substantial improvement compared to the CE-0 coupling. In other words, the ML-based interface closure is able to learn the correlation between microscale and macroscale variables solely from the data and produces more accurate prediction than the CE-0 coupling. This finding clearly illustrates the potential of ML for learning interface closure in micro-macro solvers. Although the ML-based closure yields a sufficiently accurate solution, we reiterate here that for the model problem investigated in this study, there are higher-order approximations that can be utilized to reconstruct particle distribution functions. For example, the first-order approximation of $f_1^l(x_{\text{FDM}},t)$ using the Chapman–Enskog relations (CE-1) is given below
\begin{equation}
\begin{multlined}
    f_1^u(x_{\text{FDM}},t) = \frac{u(x_{\text{FDM}},t)}{3} -  \\ \frac{\Delta x}{3 \omega^u} \frac{u(x_{\text{LBM}},t) - u(x_{\text{FDM}}-\Delta x,t)}{2 \Delta x}, 
\end{multlined}
\end{equation}
\begin{equation}
    \begin{multlined}
        f_1^v(x_{\text{FDM}},t) = \frac{v(x_{\text{FDM}},t)}{3} - \\ \frac{\Delta x}{3 \omega^v} \frac{v(x_{\text{LBM}},t) - v(x_{\text{FDM}}-\Delta x,t)}{2 \Delta x},
    \end{multlined}
\end{equation}
where the first term is the zeroth order approximation, and the second term is the first order derivative computed using the central difference scheme.  

\begin{figure}[ht]
\centering
\includegraphics[width=0.99\linewidth]{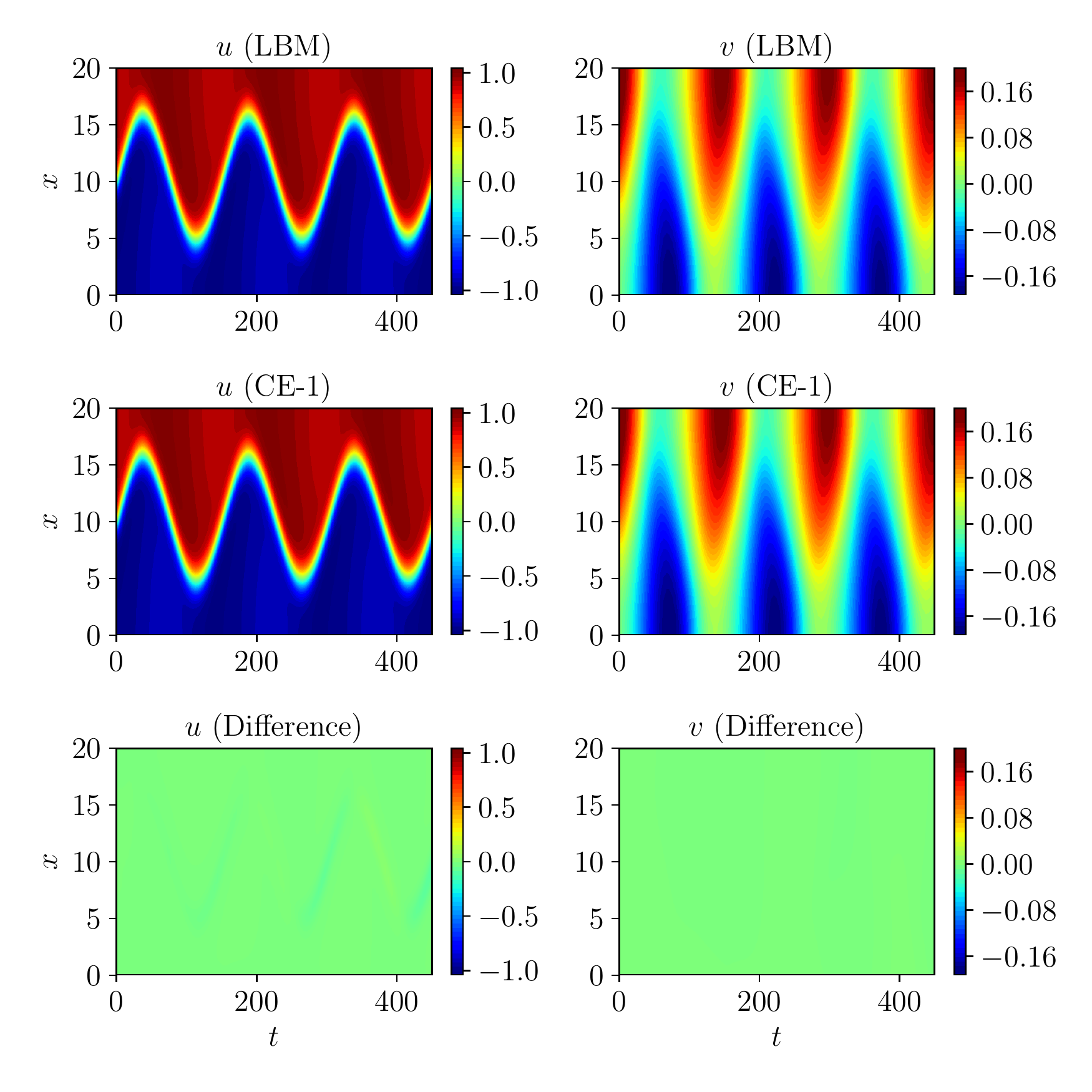}
\caption{Spatio-temporal prediction of the $u$ and $v$ field of the FitzHugh-Nagumo model by hybrid solver with CE-1 coupling at the interface.}
\label{fig:ce1}
\end{figure}

For a fair comparison, we illustrate the results of FDM-LBM solver equipped with CE-1 coupling at the interface. Figure~\ref{fig:ce1} displays the $u$ and $v$ fields for the LBM simulation, the FDM-LBM solver with CE-1 coupling at the interface, and the difference between two simulations. The agreement between LBM and hybrid solver is excellent and even better than the ML-based closure at the interface. This clearly shows limitations of data-driven methods compared to higher-order numerical approximation of analytical relations between microscale and macroscale solvers. However, obtaining such higher-order analytical approximation for complex geometries and complex local interactions might not be possible. Furthermore, for many complex systems, the evolution equations may not be available in closed forms \cite{keunings2004micro,attinger2003generalized} and for such systems, the ML-based interface closure will be attractive, especially, in the age of data.  

We assess the quantitative performance of different coupling algorithms investigated in this study using the root mean-squares error (RMSE) defined as
\begin{equation}
    \text{RMSE} = \sqrt{\frac{1}{N_t N_x} \sum_{n=1}^{N_t}\sum_{i=1}^{N_x} (\phi^{\text{T}}(x_i,t_n) - \phi^{\text{P}}(x_i,t_n))^2},
\end{equation}
where $N_t$ is the total number of time snapshots, $N_x$ is the total number of grid points, $\phi$ is the variable of interest. The superscripts $T$ and $P$ stand for the truth solution (i.e., LBM simulation) and predicted solution (i.e., FDM-LBM solver simulation). We store 450 temporal snapshots between time $t=0$ to $t=450$ for the calculation of the RMSE. Table~\ref{tab:rmse} reports the RMSE for different coupling algorithms presented in this study. The quantitative assessment implies that the error is minimum with CE-1 coupling and the error for the ML-based coupling is substantially less than the CE-0 coupling. We believe that the error for the ML-based coupling can be further reduced by including more information in the input space of the neural network or using a deeper neural network architecture. However, our experiments with hyperparameters for the neural network suggest that the deeper network leads to overfitting for this relatively simple problem and might not lead to an improvement in the prediction.  

\begin{table}[htbp!]
\caption{The root mean squared error between the true and predicted solution for $u$ and $v$ for different coupling algorithms.}
\centering
\begin{tabular}{p{0.18\textwidth} p{0.12\textwidth} p{0.12\textwidth} }  
\hline
Coupling algorithm  & $u$ & $v$  \\
\hline \\
CE-0   & $6.07 \times 10^{-1}$ & $8.01 \times 10^{-2}$ \\ 
CE-1  & $9.85 \times 10^{-3}$ & $7.49 \times 10^{-4}$ \\ 
ML  & $1.26 \times 10^{-1}$ & $1.13 \times 10^{-2}$  \\ 
\hline
\end{tabular}
\label{tab:rmse}
\end{table}



\emph{Conclusion} --- In this work, we have introduced a framework based on machine learning (ML) to learn the interface closure between microscale lattice Boltzmann method (LBM) and macroscale finite difference method (FDM) solver. This framework offers a great potential for coupled micro-macro solver where there is one to many mapping across the interface and the true relation between microscale and macroscale variables is unknown or difficult to derive, as in the case of complex geometries and multiphysics systems. The framework is modular enough that it can be implemented to different coupling models such as RANS-LES, FDM-FVM, FVM-LBM, and others. Our experiments with the FitzHugh-Nagumo model suggest that the ML-based interface closure framework is able to produce sufficiently accurate dynamics over a longer time. Also, the interface boundary closure based on the first-order approximation of Chapman–Enskog expansion (CE-1) produced even better prediction than the ML-based closure. It is no surprise that the interface closure based on the numerical approximation of the analytical relations can recover the dynamics almost exactly. However, for many multiphysics systems like viscoelastic fluids or porous media, such relations might not exist and for such systems, we envision that our framework will be more advantageous. This is the first step towards building data-driven interface closure models for micro-macro solvers, and our future efforts will be directed to extend this framework to more complex higher-dimensional multiphysics problems in fluid dynamics.

This material is based upon work supported by the U.S. Department of Energy, Office of Science, Office of Advanced Scientific Computing Research under Award Number DE-SC0019290. O.S. gratefully acknowledges their support.

The data that support the findings of this study are available from the corresponding author upon request.




\bibliography{references}
\end{document}